# Exploiting the close-to-Dirac point shift of Fermi level in Sb$_2$Te$_3$/Bi$_2$Te$_3$ topological insulator heterostructure for spin-charge conversion

E. Longo[1,†], L. Locatelli[1,†], P. Tsipas[2], A. Lintzeris[2], A. Dimoulas[2], M. Fanciulli[3], M. Longo[4], R. Mantovan[1,*]

[1]CNR-IMM, Unit of Agrate Brianza, Via C. Olivetti 2, 20864 Agrate Brianza, Italy

[2]National Centre for Scientific Research "Demokritos", Institute of Nanoscience and Nanotechnology, Agia Paraskevi, 15341, Athens, Greece

[3]Department of Material Science, University of Milano Bicocca, Via R. Cozzi 55, Milan 20125, Italy

[4]CNR-IMM, Unit of Rome, Via Fosso del Cavaliere 100, 00133 Rome, Italy

[†]Equally contributed to the work
[*]roberto.mantovan@cnr.it

**Abstract**

Properly tuning the Fermi level position in topological insulators is of vital importance to tailor their spin-polarized electronic transport and to improve the efficiency of any functional device based on them. Here we report the full *in situ* Metal Organic Chemical Vapor Deposition (MOCVD) and study of a highly crystalline Bi$_2$Te$_3$/Sb$_2$Te$_3$ topological insulator heterostructure on top of large area (4") Si(111) substrates. The bottom Sb$_2$Te$_3$ layer serves as an ideal seed layer for the growth of highly crystalline Bi$_2$Te$_3$ on top, also inducing a remarkable shift of the Fermi level to place it very close to the Dirac point, as visualized by angle-resolved photoemission spectroscopy. In order to exploit such ideal topologically-protected surface states, we fabricate the simple spin-charge converter Si(111)/Sb$_2$Te$_3$/Bi$_2$Te$_3$/Au/Co/Au and spin-charge conversion (SCC) is probed by spin pumping ferromagnetic resonance. A large SCC is measured at room temperature, which is interpreted within the inverse Edelstein effect (IEE), thus resulting in a conversion efficiency of $\lambda_{IEE} \sim 0.44 \, nm$. Our results demonstrate the successful tuning of the surface Fermi level of Bi$_2$Te$_3$ when grown on top of Sb$_2$Te$_3$ with a full *in situ* MOCVD process, which is highly interesting in view of its future technology transfer.

**Introduction**

The use of topological matter for spintronics has been attracting huge interest since it paves the way toward highly energy-efficient and ultra-fast devices.[1–3] In particular, topological insulators (TIs) represent a very interesting case due to their unconventional electronic band structure.[4,5] In principle, TIs are characterized by insulating bulk states (BS) and by linearly dispersed (i.e. Dirac-like) highly conductive topological surface states (TSS). A useful aspect of the TSS is their helical spin polarization arising from the so-called spin-momentum locking, where the spin of the electrons moving along a specific direction is locked orthogonally to this motion.[4] Recently, TIs have been proposed as promising candidates to achieve efficient data read-out in the MESO devices proposed by Intel for future processing-in-memory architectures.[1,6] TIs are of particular interest to implement spintronic devices that are based on spin-charge conversion (SCC) mechanisms.[7,8] The reciprocal effect, i.e. charge-spin conversion (CSC) has been also demonstrated at the interface between TIs and ferromagnetic (FM) layers, showing the great potentiality of TIs when compared to traditional heavy elements for spin-orbit torque-based devices. [9,10]

However, most of the 3D-TIs, such as $Bi_2Se_3$, $Bi_2Te_3$ and $Sb_2Te_3$, have their Fermi level ($E_F$) cutting the conduction and/or valence band. Therefore, the electrical conduction in TIs is often characterized by the superposition of conduction mechanisms originating from the TSS and the bulk states, the latter being detrimental in the optimization of both SCC and CSC in TIs-based devices.[11] In the specific case of $Bi_2Te_3$, few reports exist discussing its use for SCC applications and mostly conducted at low temperature.[12]

Several strategies have been proposed to engineer the position of the $E_F$ to adjust it ideally at the crossing of the linearly dispersed bands of the TSS, i.e. close to the Dirac point (DP). One of the most widely attempted strategies is to control the TIs' stoichiometry by varying the molar fraction of the constituting elements. In the case of chalcogenide-based 3D-TIs, a notable example is the $(Bi_{1-x}Sb_x)_2Te_3$ family, for which S.H. Su *et. al* [13], Okada *et al.*[14] and Kondou *et al.*[15] have demonstrated how the fine control of the $E_F$ position can be achieved by gradually changing the Sb content. In these works, the authors have showed an enhanced SCC efficiency for a specific Sb concentration, where the Fermi level is rigidly shifted in the proximity of the DP. An alternative strategy consists in controlling the chemical doping of the TIs' surfaces. In fact, D. Hsieh *et al.*[16] have demonstrated that the modification of the Ca content in $Bi_{1-x}Ca_xTe_3$ compounds can increase the surface hole donor concentration, which progressively pushes the $E_F$ position towards the DP. Similarly, the use of interlayers between TIs and neighboring FM layers can also be used to intentionally engineer the band structure of a TI, finally enhancing the SCC efficiency in TI/FM heterostructures. For instance, R. Sun et al.[17] have reported in 2019 a study on the SCC mechanisms in $Bi_2Se_3$/Bi/FM systems as a function of the Bi thickness. As a result, they have demonstrated the possibility to deeply modify the system generating a Rashba-like 2D electronic gas band structure superimposed to the TSS of a $Bi_2Se_3$ layer, hence achieving full control of the SCC efficiency of the proposed heterostructure. The direct modification of the chemistry of a material is not the only way to engineer the Fermi level position. Interestingly, Du *et al.*[18] have predicted that the internal strain present in a topological material (i.e. antiferromagnetic TI) can modify the position of the Fermi level due to the presence of defects and substitutional dopants in the structure.

In our previous work[11], we have investigated the topological properties of $Bi_2Te_3$ thin films grown by Metal Organic Chemical Vapor Deposition (MOCVD) on 4'' Si(111) substrates, demonstrating the presence of TSS lying in the bulk insulating gap of the material. Through Angle-Resolved Photoemission Spectroscopy (ARPES) measurements the Dirac-like dispersion of such surface states has been established, despite the position of the Fermi level has been found to be within the conduction band of the material, as typically observed in $Bi_2Te_3$. By performing magnetotransport experiments a clear weak anti-localization (WAL) effect, has been observed in $Bi_2Te_3$ and attributed to a 2D-type of conduction.[11] However, still a remarkable contribution from the BS to the $Bi_2Te_3$ transport has been evidenced, thus hindering the ideal contribution from the TSS. Indeed, it has already been shown by Wu *et al.*[19] that the presence of BS at the Fermi level is a competitive transport mechanism, which reduces the SCC performance of a TI, inhibiting a pure quantum topological transport through the TSS.

In the wake of these considerations, the present manuscript showcases a possible strategy to engineer the $E_F$ position in MOCVD-produced $Bi_2Te_3$ thin films. In particular, the commensurate growth of a 90 nm thick $Bi_2Te_3$ layer on top of a nearly epitaxial Si(111)/$Sb_2Te_3$ seed substrate is conducted by optimizing a full *in situ* process, where the whole heterostructure is grown by MOCVD on Si(111) over an area up to 4". The structural properties of the obtained samples are studied by X-ray diffraction (XRD) performed in Bragg-Brentano geometry, to assess their crystalline quality. Atomic force microscopy (AFM) is employed for a direct visualization of the surface morphology and the roughness estimation of the $Bi_2Te_3$ layer. ARPES measurements on the $Sb_2Te_3$/$Bi_2Te_3$ heterostructure are performed to probe the shifting of $E_F$ in the $Bi_2Te_3$ top layer, and the obtained results are compared with those previously reported on the single Bi$_2$Te$_3$ layer.[11] In order to correlate the band engineering with the functionality of the proposed heterostructure, the deposited topological bilayer is coupled with a Au(5nm)/Co(5nm)/Au(5nm) ferromagnetic trilayer grown by evaporation, and spin pumping ferromagnetic resonance (SP-FMR) measurements are performed to extract the SCC efficiency of the system. The measured SP-FMR signals are interpreted adopting the inverse Edelstein effect (IEE) model, from which a remarkable IEE length $\lambda_{IEEE} \sim 0.44\ nm$ is extracted at room temperature for the Sb$_2$Te$_3$/Bi$_2$Te$_3$ heterostructure, being a considerably high value within the class of chalcogenide-based 3D-TIs systems (see Table 1 of Ref. [20]). Here, the adoption of the Au interlayer turns out to be beneficial to protect the Bi$_2$Te$_3$ TSS from degradation, avoiding uncontrolled interfacial intermixing, as proved also for similar systems.[20–22]

**Results and Discussion**

In Figure 1(a), the XRD pattern of the Si(111)/Sb$_2$Te$_3$/Bi$_2$Te$_3$ heterostructure acquired in the Bragg-Brentano geometry is depicted. Through this measurement, the out-of-plane (OOP) orientation of the crystalline grains of the rhombohedral Bi$_2$Te$_3$ layer belonging to the R-3m space group is extracted.[23] The brighter signals emerging from the contour plot shown in Fig.1(a) with the red spots, correspond mainly to the Bi$_2$Te$_3$ crystalline planes oriented along the $[00\ell]$ direction. In order to quantify the deviation from the full OOP orientation of the $(00\ell)$ planes with respect the Si(111) surface, the mosaicity of the Bi$_2$Te$_3$ layer is calculated by extracting the rocking curve around the

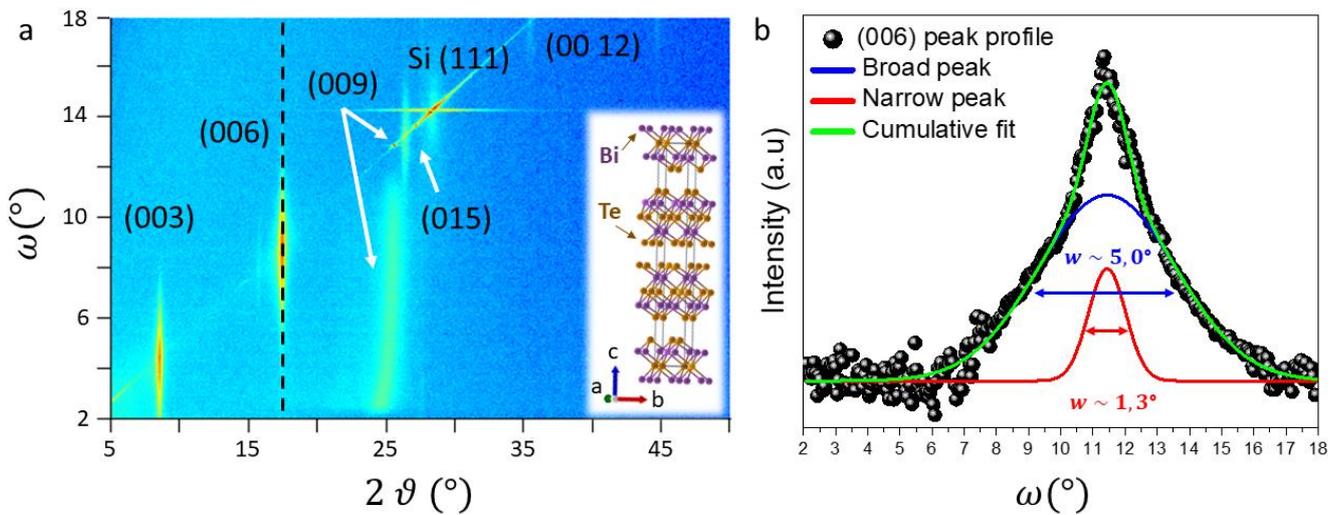

Figure 1: (a) XRD pattern acquired in the Bragg-Brentano geometry. The more intense signals (red elongated dots) indicate the OOP orientation of the Bi$_2$Te$_3$ crystalline planes (inset: Bi$_2$Te$_3$ crystalline structure according to ICSD code 74348). (b) Rocking angle profile of the (006) reflection. The acquired data is fitted with different Gaussian curves to extract the value of the mosaicity for the different families of oriented crystals.

(006) reflection (black dashed line in Fig.1(a)). In Fig.1(b) the variation of the signal intensity as a function of the rocking angle ω is reported and fitted with a Gaussian curve (green solid line). As a first Gaussian fitting attempt, a single peak was employed, which does not provide a satisfactory fitting of the collected data. Therefore, an alternative fitting strategy is employed by considering contributions coming from the presence of several families of crystalline grains, being characterized by a different mosaicity. The result is shown in Fig. 1(b), where a very good fit of the data (green solid line) is achieved following the introduction of two Gaussian components: a broad (blue solid line) and a narrow one (red solid line), corresponding to a mosaicity of $(5,00 \pm 0,08)°$ and $(1.30 \pm 0,06)°$, respectively. Although the more intense signals correspond to the $(00\ell)$ planes, additional reflections can be found on the Bragg-Brentano line (i.e. $\omega = 2\theta$), indicating that some $Bi_2Te_3$ crystals are OOP oriented along different directions (i.e. [015] as shown in Fig.1(a)).

The analysis of the Bragg-Brentano diffraction pattern evidences a high degree of crystallinity of the $Bi_2Te_3$ film. However, the limited but still elongated shape of the Bragg-Brentano signals suggests the presence of two families of crystalline grains with a different mosaicity and the elongated reflection at $2\theta = 26°$ indicates that some disorder is still present in the film. To provide a complete description of the residual disorder in the grown $Bi_2Te_3$ films, a grazing incidence X-ray diffraction (GIXRD) measurement is performed and reported in Fig. S1 of the Supporting Information. Being the GIXRD a volume measurement, it can be considered as a probe of the overall chemical-structural quality of a crystalline film. In this case, in the GIXRD pattern few reflections are present that do not belong to the $(00\ell)$ family of planes, thus confirming the presence of a polycrystalline fraction within the $Bi_2Te_3$ layer thickness.

To provide further insights into the surface morphology of the $Bi_2Te_3$ layer, AFM measurements are conducted on different areas of the sample. In Figure 2(a) an AFM acquisition on a 30x30 µm$^2$ area of the $Bi_2Te_3$ bare surface is reported. Here, several in-plane oriented $Bi_2Te_3$ hexagonal-like flakes are clearly visible, confirming the high crystalline nature of the deposited film, its crystalline symmetry, and a remarkable in-plane order.[23] From the same picture, it turns out that some grains are merged in a more disordered fashion, a fact which does not always allows to distinguish the shape of the single flake. This evidence is fully in accordance with the XRD analysis discussed above, confirming the coexistence of highly texturized $Bi_2Te_3$ regions with more disordered polycrystalline portions.

In Fig. 2(b) an AFM scan on a 5x5 µm² smaller area is performed to highlight the flake geometry and to precisely extract its height profile, as from the black solid square in panel (a). The right side of Fig.2 (b) shows the linear profile corresponding to the black dashed line in the left picture of panel (b), from which it turns out that the height of the selected flake is approximately 20 nm, thus being composed by about 20 Bi$_2$Te$_3$ quintuple layers (QLs). To have an estimation of the surface roughness within a single Bi$_2$Te$_3$ flake, an additional scan is performed on a 0.5x0.5 µm² area, as reported in Fig. 2(c). From the analysis of the height profile of this portion of the surface, a very smooth morphology is evidenced, being characterized by a main square roughness (RMS, Rq) of 0.69 nm, which is slightly higher than the one calculated for the bare Si(111)/Bi$_2$Te$_3$ reported in Ref.[23] for similar MOCVD deposited films, namely 0.5 nm.

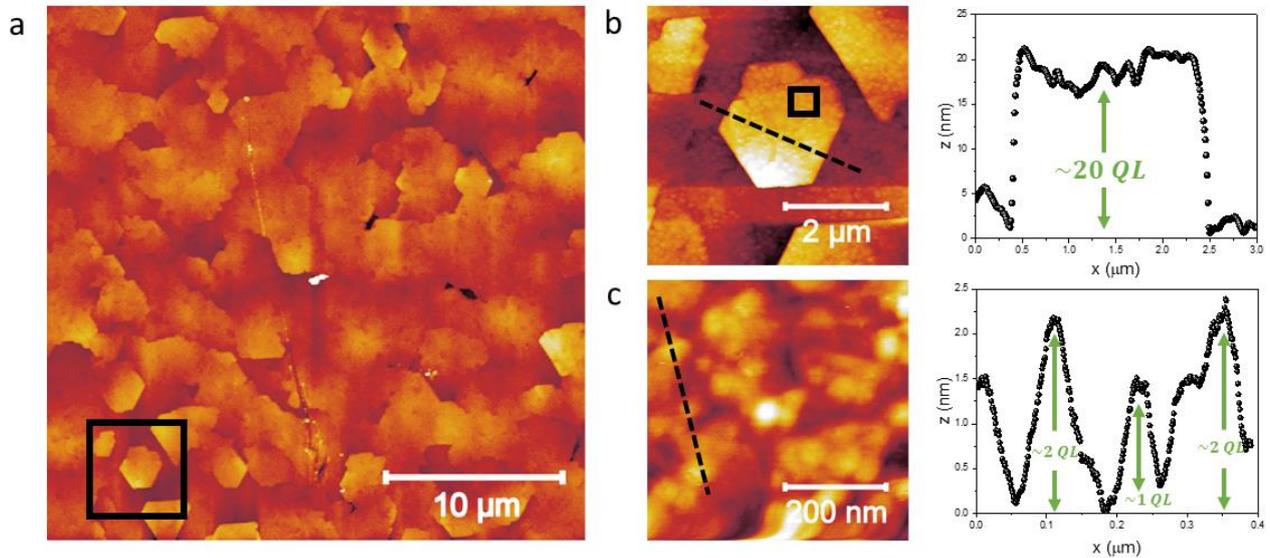

Figure 2: AFM measurement of the Bi$_2$Te$_3$ uncapped surface. (a) Scan area of 30x30 µm², where the presence of several triangular like flakes are clearly visible. In the inset a 5x5 µm² scan of the area highlighted by the dashed black square in (a) is reported. The height profile of the Bi$_2$Te$_3$ flake is reported in panel (b). In (c) a scan limited to a 0.5x0.5 µm² area is show. From panel (c) we measured a morphological RMS roughness of 0.69 nm.

ARPES technique is ideal to conduct the direct visualization of the band structure of the Bi$_2$Te$_3$'s close-to-the surface layers. Figure 3 shows a direct comparison between the band structure of Bi$_2$Te$_3$ when grown directly on Si(111) (Fig.3(a)),[11] and on top of Sb$_2$Te$_3$ (Fig.3(b)). Being ARPES extremely sensitive to the first few surface layers, a cleaning procedure was conducted prior to measurements, in order to remove the surface contaminants (see Methods and Supp. Info. for details).

As it emerges by the ARPES image reported in Fig. 3(a), in the bare Si(111)/Bi$_2$Te$_3$ heterostructure the Fermi level is clearly positioned across the conduction band of the material, influencing the overall electronic transport, as revealed by magnetotransport measurements reported in Ref. [11]. On the other hand, by comparing this result with the panel (b) of the same figure, the Fermi level is rigidly shifted within the bulk insulating gap when the $Sb_2Te_3$ seed layer is placed between the Si substrate and $Bi_2Te_3$, intercepting the TSS in the proximity of the DP.

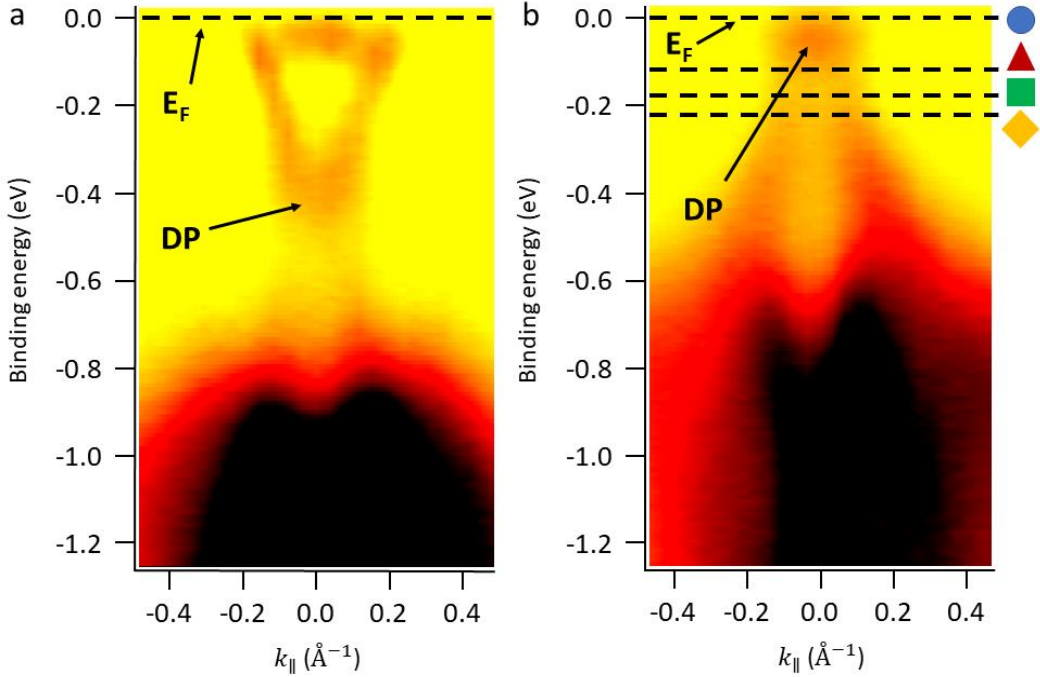

Figure 3: ARPES intensity map obtained for Si(111)/$Bi_2Te_3$, in panel (a), and Si(111)/$Sb_2Te_3$/$Bi_2Te_3$ heterostructures, in panel (b). The black dashed lines in panel (b) indicate the different energies at which the contour plot is reported in Fig.4.

By comparing the two panels of Fig. 3 it can be noticed that the $Bi_2Te_3$ grown on top of the $Sb_2Te_3$ seed layer shows a rigid shift of the band structure with respect to the layer grown on top of Si(111). A similar observation has been reported by V. Pereira *et al.*[24] for the reversed stack: $Bi_2Te_3$(bottom)/$Sb_2Te_3$(top), where the $Sb_2Te_3$ layer is subject to a modification of its band structure when just a few QLs are deposited, with a rigid shift of the chemical potential when a thicker layer is deposited. The latter condition is attributed to the different in-plane lattice constants of $Sb_2Te_3$ and $Bi_2Te_3$, being 0.42 nm and 0.44 nm, respectively. Such a 5% difference could induce an effective compressive (tensile) strain when $Bi_2Te_3$ is on top (bottom), to generate the observed chemical potential shift.

Different ARPES polar maps are acquired in this work at room temperature for the Si(111)/$Sb_2Te_3$/$Bi_2Te_3$ system at various binding energies, with the aim to follow the evolution of the band structure from the surface to the bulk states of the $Bi_2Te_3$ top layer. In Figure 4 we report the ARPES contour plots for E = 0 (Fermi level) eV, -0.12 eV, -0.18 eV and -0.22 eV, as indicated by the colored symbols marked in panel (b) of Fig. 3.

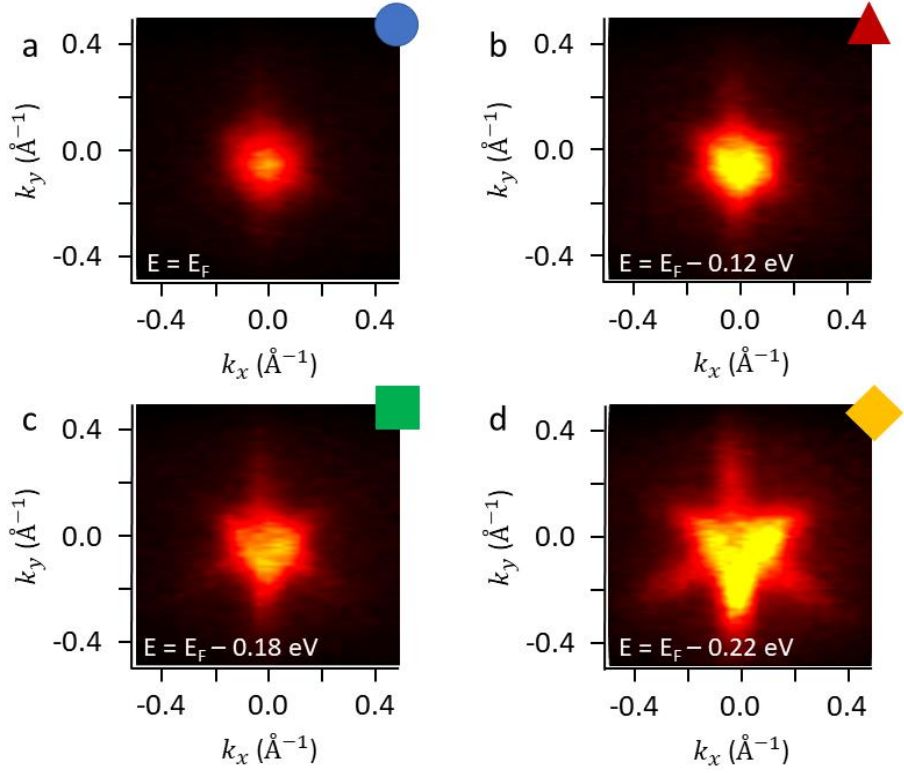

Figure 4: ARPES energy contour plots for the Si(111)/Sb$_2$Te$_3$/Bi$_2$Te$_3$ heterostructure. The K$_x$ and K$_y$ components were acquired along the K-G-K and M-G-M crystallographic direction in the k-space. The colored symbols indicate the various energy levels as indicated in Fig. 2 (b).

The pointwise symmetry of the contour plot in Fig. 4(a) for E = $E_F$ indicates that the system lies in the proximity of the DP, a condition which highlights the 2D quantum transport properties of the Si(111)/Sb$_2$Te$_3$/Bi$_2$Te$_3$ heterostructure. For energies lower than $E_F$, the shape of the spectra assumes different geometries. At E = - 0.12 eV, the hexagonal pattern appears, as expected for the symmetry rules of the TSS. Pushing down to E = -0.18 eV a superposition between a hexagon and a triangular shape is observed, as a clear indication that the TSS and the BS are co-present at this energy. Finally, at E = - 0.22 eV the ARPES pattern is fully trigonal, therefore showing that the energy level is now solely crossing the Bi$_2$Te$_3$ BS.

Having successfully moved the $E_F$ close to the DP, it is now highly of interest to exploit the Bi$_2$Te$_3$ functionality for SCC by making use of SP-FMR. In SP-FMR pure spin currents are generated into the FM layer, and subsequently converted into charge currents flowing across the surface states of the TI.[20,25] However, it is known that the direct contact between a FM with a TI can be detrimental for the TSS,[15,20,22] with the adoption of appropriate non-magnetic and low spin-orbit coupling interlayers often being beneficial to preserve them and avoid chemical intermixing to successfully exploit SCC.[20–22] We follow the same methodology as previously employed to probe SCC into Sb$_2$Te$_3$.[20,22] In particular, we use Co as FM and Au as inter-/capping- layers, with the following final stack structure (from the bottom): Si(111)/Sb$_2$Te$_3$/Bi$_2$Te$_3$/Au(5nm)/Co(5nm)/Au(5nm). To properly estimate the SCC efficiency, the signals extracted by the functional heterostructure comprising the TI material are compared with a Si(111)/Au(5nm)/Co(5nm)/Au(5nm) reference heterostructure prepared simultaneously, during the same Au/Co/Au evaporation process. In the following, the Si(111)/Au/Co/Au and Si(111)/Sb$_2$Te$_3$/Bi$_2$Te$_3$/Au/Co/Au heterostructures are named with S0 and S1, respectively.

To extract the magnetization dynamics parameters necessary to quantify the SCC efficiency, broadband ferromagnetic resonance (BFMR) measurements are performed.[20,26] In Figure 4(a) the Kittel curves acquired for heterostructures S0 and S1 are reported (black squares and red circles, respectively), where the resonant frequency ($f_{res}$) is plotted as a function of the resonant magnetic field ($H_{res}$) (see Methods). The evolution of the $f_{res}(H_{res})$ response is acquired over a large frequency range (11-30 GHz), to ensure a reliable quantification of the parameters extracted from the fit with Equation 1 (red solid line). For a polycrystalline FM thin film positioned in the in-plane (IP) configuration, the Kittel equation can be expressed as:

$$f_{res} = \frac{\gamma}{2\pi} \sqrt{H_{res}\left(H_{res} + 4\pi M_{eff}\right)} \qquad (1)$$

where $\gamma$ is the gyromagnetic ratio and $M_{eff}$ the effective magnetization. $\gamma = g \frac{e}{2m_e} \left[\frac{Hz}{Oe}\right]$, where $e$ and $m_e$ are the charge and the mass of the electron, and $g$ is the Landè g-factor, a quantity which links the electronic angular and spin momenta of a material.[27]

After extracting the $M_{eff}$ value from Eq.1, some information on the magnetic anisotropy can be achieved by writing the relation $4\pi M_{eff} = 4\pi M_s - H_k = 4\pi M_s - \frac{2K_s}{M_s t_{Co}}$, where $M_s$, $t_{Co}$, $H_k$ and $K_s$ are the saturation magnetization, the thickness of the Co layer, the magnetic anisotropy field and the surface magnetic anisotropy constant, respectively.

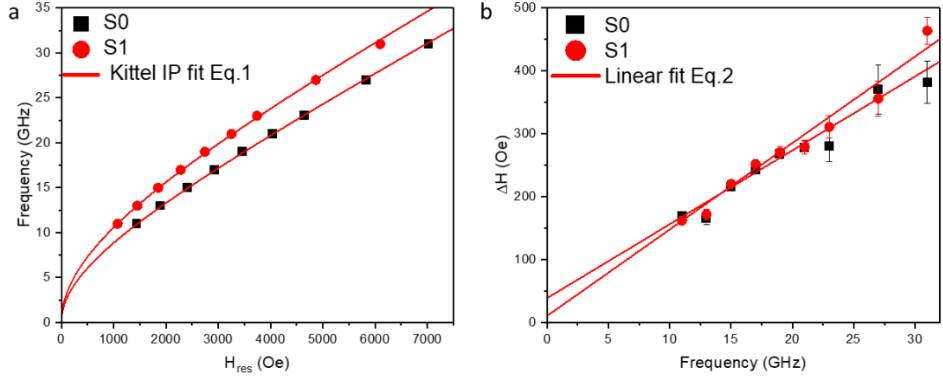

Figure 5: (a) Kittel dispersion for the IP configuration. The dataset for heterostructures S0 (black squares) and S1 (red circles) are reported and fitted with Eq. 1 (red solid line). (b) The signal linewidth $\Delta H$ as a function of the resonant frequency $f_{res}$ is reported for the same heterostructures of panel (a) and fitted with Eq.2 (red solid line).

Adopting the same methodology of Refs.[20,22], from the fit of the Kittel dispersion reported in Figure 1(a) we obtain $\gamma^{S0} = (1.91 \pm 0.01) \cdot 10^7 \frac{Hz}{Oe}$ ($g^{S0} = 2.17 \pm 0.01$) and $\gamma^{S1} = (1.85 \pm 0.04) \cdot 10^7 \frac{Hz}{Oe}$ ($g^{S0} = 2.15 \pm 0.05$). The extracted $g$ values are very similar and perfectly in agreement with most of those found in literature ($g \sim 2.25$). The corresponding effective magnetizations are $M_{eff}^{S0} = 609 \pm 9 \frac{emu}{cm^3}$ and $M_{eff}^{S1} = 973 \pm 59 \frac{emu}{cm^3}$, for S0 and S1, respectively. These values are lower than the Co bulk value ($\sim 1400 \frac{emu}{cm^3}$)[28], as typically observed in thin films as due to the possible presence of magnetic dead layers and/or to magnetic shape anisotropy.[29]

In Figure 5(b), the linewidth $\Delta H$ of the BFMR signal is plotted as a function of $f_{res}$ and fitted according to the following Eq. 2.

$$\Delta H = \Delta H_0 + \frac{4\pi}{|\gamma|} \alpha f_{res} \qquad (2)$$

Here, $\alpha$ represents the damping constant and $\Delta H_0$ the inhomogeneous broadening. From the FMR theory,[27,30] $\alpha$ accounts for how fast the magnetization vector $\boldsymbol{M}$ in a FM aligns along an applied external magnetic field at the resonance condition.[27] Here, according to the linear fit (red solid lines) of the datasets reported in Figure 5(b), $\alpha^{S0} = (17.9 \pm 1.3) \cdot 10^{-3}$, $\Delta H_0^{S0} = 38 \pm 13\ Oe$, $\alpha^{S1} = (20.3 \pm 1.2) \cdot 10^{-3}$, and $\Delta H_0^{S1} = 10 \pm 12\ Oe$ are extracted. $\Delta H_0$ provides information on the magneto-structural properties of the FM layer, accounting for samples imperfections or texturing.

This value is acceptable ad comparable for both samples, indicating a limited amount of magnetic disorder in the Co layer (i.e. magnetic dead layers, structural imperfections, etc.).

According to the spin pumping theory[25,31], the difference between the $\alpha$ values calculated for S0 and S1 is proportional to the real part of the spin mixing conductance $Re(g_{eff}^{\uparrow\downarrow})$ characterizing the interface between the FM and the adjacent spin-sink layer, in this case the $Bi_2Te_3$ layer. $Re(g_{eff}^{\uparrow\downarrow})$ is an intrinsic quantity of a system, which accounts for the pure spin current flowing across the interface towards the $Bi_2Te_3$. In our case, by considering the measured $\Delta\alpha = (2.4 \pm 2.5) \cdot 10^{-3}$, we obtain $Re(g_{eff,Bi2Te3}^{\uparrow\downarrow}) = 7,53 \cdot 10^{18} \, m^{-2}$, a value perfectly in agreement with those extracted elsewhere for similar systems (see Table 1 Ref.[20]). For the details about the procedure to extract the latter value, see the Supplementary Information.

The enhancement of $\alpha$ alone is a necessary but not sufficient evidence to conclude about the existence of SCC, which can only be demonstrated by electrically detecting the spin pumping signal. In Figure 6, the electrically detected SP-FMR measurements carried out on the S0 and S1 heterostructures are shown. A detailed description of the experimental procedure and the theoretical background can be found in Refs.[22,31,32]

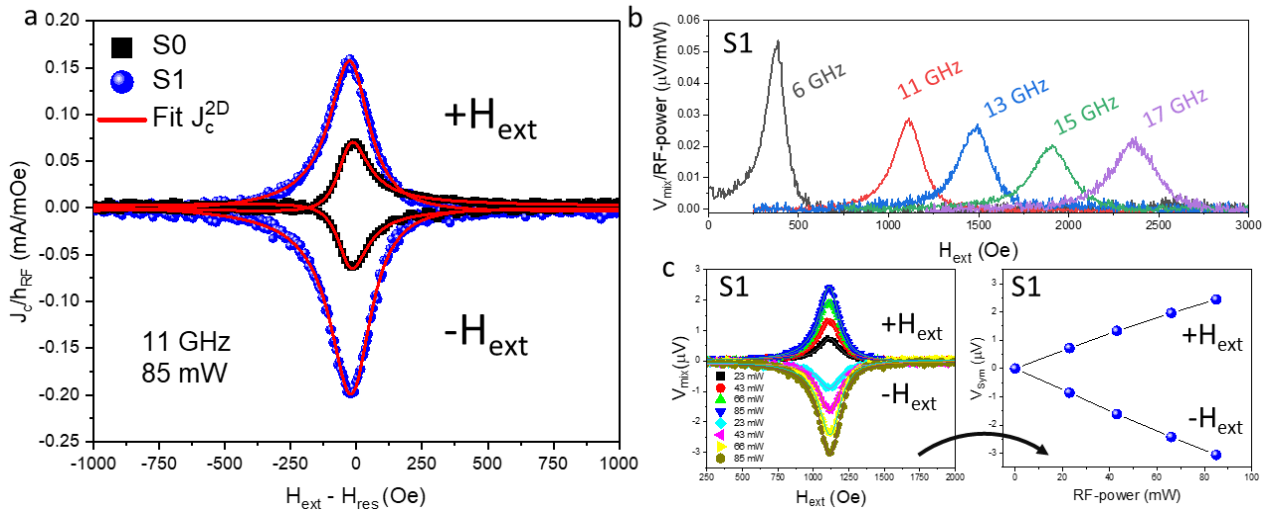

Figure 6: Electrically-detected SP-FMR measurements. (a) 2D charge current generated in resonant condition and normalized to the strength of the oscillating magnetic field ($h_{RF}$) for heterostructure S0 (black squares) and S1 (blue circles) at a frequency of 11 GHz and 85 mW of RF power. The red solid line indicates the fit with Eq.3. (b) Frequency-dependent response for heterostructure S1. (c) (left) $V_{mix}$ vs $H_{ext}$ curves as a function of the RF-power acquired at positive and negative applied external magnetic fields for heterostructure S1. (right) Dependence of the voltage symmetric component $V_{Sym}$ as a function of the RF-power for the curves reported in the left panel.

In Figure 6(a) the mixing charge current density $J_{mix}$ is plotted for the heterostructure S0 (black squares) and S1 (blue circles), where the red solid lines indicate the fit of both curves, as obtained by the following Lorentzian function of Eq. (3).

$$\frac{J_{mix}}{h_{RF}} = \frac{V_{mix}}{WRh_{RF}} = \frac{1}{WRh_{RF}}\left[V_{Sym}\frac{\Delta H^2}{\Delta H^2+(H-H_{res})^2} + V_{Asym}\frac{\Delta H(H-H_{res})}{\Delta H^2+(H-H_{res})^2}\right] \quad [3]$$

where $h_{RF}$, $W$ and $R$ represent the transverse oscillating magnetic field, the width of the heterostructure and its sheet resistance, respectively. In this case the adopted values are $h_{RF} = 0.73\ Oe$, $W_{S0} = 2\ mm$, $W_{S1} = 2\ mm$, $R_{S0} = 16\ Ohm$ and $R_{S1} = 10\ Ohm$ (measured with a four-point probe). From such a fit, it is possible to isolate the symmetric component of the curve, which is proportional to the so-called symmetric voltage $V_{Sym}$ and directly connected to the spin pumping taking place at the TI/FM interface. Differently, the asymmetric part ($\propto V_{Asym}$) depends on spurious rectification effects (i.e. Oersted field, anisotropic magnetoresistance, etc.), thus to be excluded from the estimation of the SCC mechanism.[33,34] The curves in Fig. 6(a) are highly symmetric, thus indicating that the rectification effects are not the main processes driving the electrical response in both heterostructures. Despite the similar shape of these curves, the intensity of the electrical signals generated from S1 is much higher than that in S0, at the same RF power, suggesting that the insertion of the $Sb_2Te_3/Bi_2Te_3$ bilayer is beneficial to generate a higher charge current density in S1, when compared to the reference S0. In order to remove a possible thermal component (i.e. Seebeck effect[35]) from the symmetric part of $J_{mix}$, the actual signal arising from SP is obtained mediating the symmetric part of the $J_{mix}$ curves acquired for positive and negative magnetic fields, according to the equation $\frac{J_{SP}}{h_{RF}} = \frac{1}{h_{RF}}\left[\frac{J_{Sym}(+H_{ext})-J_{Asym}(-H_{ext})}{2}\right]$. The extracted values are $\frac{J_{SP}^{s0}}{h_{RF}} = 0.064 \pm 0.005\ mA/Oe$, $\frac{J_{SP}^{s1}}{h_{RF}} = 0.180 \pm 0.001\ mA/Oe$, $\frac{J_{Asym}^{s0}}{h_{RF}} = 0.027 \pm 0.006\ mA/Oe$ and $\frac{J_{Asym}^{s1}}{h_{RF}} = 0.024 \pm 0.002\ mA/Oe$. For a quantitative estimate of the extra-current produced in S1 when compared to S0, we underline here the ratio $\frac{J_{SP}^{s1}}{J_{SP}^{s0}} = 2.84 \pm 0.24$, representing a clear and direct evidence for the role played by the insertion of the $Sb_2Te_3/Bi_2Te_3$ heterostructure in generating remarkable SCC. As a further confirmation that the SCC mechanism observed in S1 (Fig. 6(a)) is consistent with the spin pumping effect, in Fig. 6 (b) the SP-FMR signals acquired for different RF frequencies are reported, demonstrating the relation between the measured voltage and the BFMR characteristic of the system. Moreover, the $V_{Sym}$ component of the IEE signal should be linear when plotted as a function of the RF power, as showed in Fig. 6 (c).[36]

In a spin pumping experiment, the SCC efficiency can be estimated mainly according to two models: the Inverse Spin-Hall effect (ISHE) and the Inverse Edelstein effect (IEE). For the ISHE the spin current generated into the FM layer is pumped and converted in the bulk states of the spin-sink layer (i.e. $Bi_2Te_3$), differently from the IEE, where solely the TSS are involved in the SCC. In our case, we perform several trials to measure the SCC in the interlayer-free $Bi_2Te_3$/Co heterostructure, but no significant SP is observed with respect to the reference heterostructure, as for the case of the $Sb_2Te_3$ TI previously investigated by some of us.[20,22] Few works exist reporting the SCC effects in $Bi_2Te_3$ films directly in contact with a FM layer. For instance, in the very recent work of X. Liu et al.[12], the authors have observed an efficient spin current injection in the $Bi_2Te_3$/MnTe heterostructure only at T<25 K, with a significant reduction around 100 K. Conversely, the introduction of interlayers at the $Bi_2Te_3$/FM interface is a reliable strategy to improve the SCC efficiency, as demonstrated by C. Li et al.[37] adopting a BN/$Al_2O_3$ tunnelling barrier.

The most plausible scenario in our case is that the Au interlayer is beneficial in protecting the $Bi_2Te_3$'s TSS detected by ARPES (Fig.3), similarly what has occurred in $Sb_2Te_3$.[20,22] Therefore, the observed SCC can be represented in terms of the IEE, where the IEE diffusion length $\lambda_{IEE} = \frac{J_{SP}}{J_s^{3D}}$ [nm] is adopted as a figure of merit to determine the conversion efficiency, with $J_s^{3D}$ being the spin current density generated in the system (see Supporting Information). According to the parameters reported in Fig. 6 (a) and to the FMR response in Fig. 5, we have $J_S^{3D} = 2.97 \cdot 10^5 Am^{-2}$, and therefore $\lambda_{IEE} = \frac{J_{SP}}{J_s^{3D}} \sim \frac{J_{Sym} h_{RF}}{J_s^{3D}} \sim \frac{0.13}{2.97 \cdot 10^5 Am^{-2}} \sim 0.44\ nm$. Despite our evidence is consistent with the IEE, the origin of the voltage signals in SP-FMR experiments is often highly debated (see Supplementary Information).[38–40] For the sake of clarity, a different fitting approach can be performed by fixing $g \sim 2.25$ in Eq. 1 from the averaging of some values found in literature for Co thin films. In this case $\lambda_{IEE}$ would be $\sim 0.40\ nm$, in strong accordance with the value reported above.

By comparing the $\lambda_{IEE}$ extracted here with some of those found in the literature, we can infer that the 0.44 nm value calculated in our MOCVD deposited heterostructure is a very large one (see Table 1 Ref.[20]). If we refer only to the SCC efficiency extracted from different chalcogenide-based topological compounds, it emerges how this value is among the highest so far reported at room temperature. Notable TI-based systems to compare with are, for example, those reported by Mendes et al.[41] with $\lambda_{IEE} = 0.075\ nm$ in $(Bi_{0.22}Sb_{0.7})_2Te_3$, or $\lambda_{IEE}$= 0.28 nm in $Bi_2Se_3$ shown by Sun et al.[17], and recent works by some of us[20,22], showing that $\lambda_{IEE}$= 0.28 nm for a 30 nm thick $Sb_2Te_3$ epitaxial TI. Also, among the possible different interlayers, the choice of Au seems to be particularly favorable. As a comparison, in the work from H. He *et al.*[42], published in 2021, the authors exploited Ru and Ti interlayers in $Bi_2Te_3$/interlayer/CoFeB systems, extracting $\lambda_{IEE}$ values ten times smaller than those we extract here.

**Conclusions**

The need to properly tune the Fermi level position in topological insulators is of paramount importance to fully make use of their topologically-protected, spin-polarized electronic transport. In this work, we report a full *in situ* MOCVD process where highly crystalline 90 nm thick $Bi_2Te_3$ thin films are grown on top of epitaxial $Sb_2Te_3$ over 4" Si(111) substrates. The use of $Sb_2Te_3$ as a seed layer, turns out to be a valid strategy to shift the $Bi_2Te_3$'s Fermi level from the conduction band (as it is in single $Bi_2Te_3$ layer) to the insulating gap in very close proximity to the Dirac point, as clearly visualized by ARPES measurements. This generates ideally pure TSS, which we exploit in a simple spin-charge converter by conducting SP-FMR measurements. A remarkably high SCC efficiency is measured in the Si(111)/$Sb_2Te_3$/$Bi_2Te_3$/Au/Co/Au heterostructure, and by interpreting such a conversion within the inverse Edelstein effect, we extract a high conversion efficiency of $\lambda_{IEE} \sim 0.44 \, nm$ at room temperature. Here, we demonstrate that the full *in situ* MOCVD of the $Bi_2Te_3$/$Sb_2Te_3$ heterostrucure is successful in producing a topological insulator characterized by ideal topologically-protected surface states, and those states can directly be exploited for efficient spin-charge conversion. This work paves the way for the future adoption of chemically deposited $Bi_2Te_3$ thin films on a large area Si wafer, thanks to the highly efficient SCC achieved via Fermi level engineering.

**Materials and Methods**

A 90 nm thick $Bi_2Te_3$ layer is grown on a 4'' Si(111)/$Sb_2Te_3$(30 nm) wafer by means of Metal Organic Chemical Vapor Deposition (MOCVD). Prior to the deposition of $Sb_2Te_3$, the Si(111) substrate is treated by means of HF acid to remove the native oxide and an annealing process is performed to properly reconstruct the Si surface. Following the substrate conditioning, growth of the seed layer of $Sb_2Te_3$ is performed at room temperature, for 90 minutes and at a pressure of 15 mbar. The deposition is then followed by a post-growth annealing, needed to promote further crystallization.

Without removing the sample from the reactor, the chamber is brought to 350° C and the deposition of the layer of $Bi_2Te_3$ is performed at 75 mbar, for 190 minutes. The growth parameters of the two TI layers are kept unchanged with respect to those employed to grow the $Sb_2Te_3$ and $Bi_2Te_3$ single layers on top of Si(111), as reported in Refs.[23,43].

The Si(111)/$Sb_2Te_3$/$Bi_2Te_3$ heterostructures dedicated to the SP-FMR studies are cut in smaller pieces and transferred into an Edwards Auto306 e-beam evaporation tool, together with the Si(111) reference heterostructures. A simultaneous deposition of the Au(5nm)/Co(5nm)/Au(5nm) trilayer is conducted for all the heterostructures.

The Si(111) substrates used as references are cleaned with isopropyl alcohol and treated with HF prior to the evaporation processes.

The Bragg-Brentano X-ray Diffraction (XRD) pattern is acquired with a HRXRD IS2000 diffractometer equipped with a Cu $K_\alpha$ radiation source ($\lambda$ = 1.5406 Å), a four-circle goniometer, and a curved 120° position-sensitive detector (Inel CPS-120). This configuration allows the detection of the asymmetric reflections produced by the crystalline planes not perfectly parallel to the sample surface, giving access to the value of the mosaicity of the crystalline grains composing the material.

Atomic Force Microscopy (AFM) images are obtained using a Bruker Dimension Edge instrument operating in tapping mode and using a sharp silicon AFM probe (TESPA, Bruker) with a typical radius of curvature in the 8–12 nm range. A polynomial background correction is applied to the raw data. To quantify the morphological surface, the Root Mean Square roughness (RMS roughness, $R_q$) value is adopted and expressed in nanometers. The different AFM images reported in the main text represent three independent measurements acquired on the areas 30x30 $\mu m^2$, 5x5 $\mu m^2$ and 0.5x0.5 $\mu m^2$.

ARPES spectra are acquired at room temperature with a 100 mm hemispherical electron analyzer equipped with a 2D CCD detector (SPECS). The He I (21.22 eV) resonant line is employed to excite photoelectrons, yielding an energy resolution of 40 meV. The spot of the employed ARPES facility has an elliptical shape with an area of about 4 × 6 $mm^2$. Thanks to this, the ARPES characterization conducted on our samples makes it possible to extract information on the dispersion of the materials band structure on a relatively macroscopic area. ARPES is performed *ex-situ*, thus a two steps surface cleaning procedure is needed to remove the oxidized species and contaminants, namely surface ion sputtering (Cr+ at 1.5 KeV) and annealing treatment at 270°C . Subsequently, the film surface is probed with Reflection High-Energy Electron Diffraction and X-ray photoemission spectroscopy to verify the effectiveness of the treatments (see Supplementary Information).

BFMR is performed using a broadband Anritsu-MG3694C power source (1-40 GHz), connected to a grounded coplanar waveguide, where the samples are mounted in a flip-chip configuration (the FM film is located close to the GCPW surface), with a 75 μm thick Kapton foil stacked in between to prevent the shortening of the conduction line. The sample-GCPW system is positioned between the polar extensions of a Bruker ER-200 electromagnet maintaining its surface parallel to the external magnetic field $H_{ext}$, in the so called in-plane (IP) configuration. During the measurements, an RF current at a fixed frequency is carried toward the GCPW and the transmitted signal is directed to a rectifying diode, converting the RF-signal in a continuous DC-current, subsequently detected by a lock-in amplifier downwards the electronic line. The same instrumentation is adopted to conduct SP-FMR measurements. Here, the edges of the sample are contacted with Ag paint and connected to a nanovoltmeter. A DC-voltage is detected in resonant condition, fixing the RF frequency and power.

**Acknowledgement**

We acknowledge financial supports from the Horizon 2020 project SKYTOP "Skyrmion-Topological Insulator and Weyl Semimetal Technology" (FETPROACT-2018-01, n. 824123), and the PNRR MUR project PE0000023-NQSTI.

**Bibliography**

1. Guo, Z. *et al.* Spintronics for Energy- Efficient Computing: An Overview and Outlook. *Proceedings of the IEEE* **109**, 1398–1417 (2021).

2. He, M., Sun, H. & He, Q. L. Topological insulator: Spintronics and quantum computations. *Frontiers of Physics 2019 14:4* **14**, 1–16 (2019).

3. Sinova, J. & Žutić, I. New moves of the spintronics tango. *Nat Mater* **11**, 368–371 (2012).

4. Hasan, M. Z. & Kane, C. L. Colloquium: Topological insulators. *Rev Mod Phys* **82**, 3045–3067 (2010).

5. Moore, J. E. The birth of topological insulators. *Nature* vol. 464 194–198 Preprint at https://doi.org/10.1038/nature08916 (2010).

6. Puebla, J., Kim, J., Kondou, K. & Otani, Y. Spintronic devices for energy-efficient data storage and energy harvesting. *Commun Mater* **1**, 1–9 (2020).

7. Soumyanarayanan, A., Reyren, N., Fert, A. & Panagopoulos, C. Emergent phenomena induced by spin-orbit coupling at surfaces and interfaces. *Nature* vol. 539 509–517 Preprint at https://doi.org/10.1038/nature19820 (2016).

8. Barla, P., Kumar Joshi, V., Somashekara Bhat, · & Joshi, V. K. Spintronic devices: a promising alternative to CMOS devices. *J Comput Electron* **20**, 805–837 (2021).

9. Wang, Y. *et al.* Room temperature magnetization switching in topological insulator-ferromagnet heterostructures by spin-orbit torques. *Nat Commun* **8**, (2017).

10. Mellnik, A. R. *et al.* Spin-transfer torque generated by a topological insulator. *Nature* **511**, 449–451 (2014).

11. Locatelli, L. *et al.* Magnetotransport and ARPES studies of the topological insulators $Sb_2Te_3$ and $Bi_2Te_3$ grown by MOCVD on large-area Si substrates. *Sci Rep* **12**, (2022).

12. Liu, X. *et al.* Temperature dependence of spin - orbit torque-driven magnetization switching in in situ grown $Bi_2Te_3$/MnTe heterostructures. *Appl Phys Lett* **118**, (2021).

13. Su, S. H. *et al.* Spin-to-Charge Conversion Manipulated by Fine-Tuning the Fermi Level of Topological Insulator $(Bi_{1-x}Sb_x)_2Te_3$. *ACS Appl Electron Mater* **3**, 2988–2994 (2021).

14. Okada, K. N. *et al.* Enhanced photogalvanic current in topological insulators via Fermi energy tuning. *Phys Rev B* **93**, 081403 (2016).

15. Kondou, K. *et al.* Fermi-level-dependent charge-to-spin current conversion by Dirac surface states of topological insulators. *Nat Phys* **12**, 1027–1031 (2016).

16. Hsieh, D. *et al.* A tunable topological insulator in the spin helical Dirac transport regime. *Nature 2009 460:7259* **460**, 1101–1105 (2009).

17. Sun, R. *et al.* Large Tunable Spin-to-Charge Conversion Induced by Hybrid Rashba and Dirac Surface States in Topological Insulator Heterostructures. *Nano Lett* **19**, 4420–4426 (2019).

18. Du, M. H., Yan, J., Cooper, V. R. & Eisenbach, M. Tuning Fermi Levels in Intrinsic Antiferromagnetic Topological Insulators $MnBi_2Te_4$ and $MnBi_4Te_7$ by Defect Engineering and Chemical Doping. *Adv Funct Mater* **31**, (2021).

19. Wu, H. *et al.* Room-Temperature Spin-Orbit Torque from Topological Surface States. *Phys Rev Lett* **123**, 207205 (2019).

20. Longo, E. *et al.* Large Spin-to-Charge Conversion at Room Temperature in Extended Epitaxial $Sb_2Te_3$ Topological Insulator Chemically Grown on Silicon. *Adv Funct Mater* **2109361**, (2021).

21. Galceran, R. *et al.* Passivation of $Bi_2Te_3$ Topological Insulator by Transferred CVD-Graphene: Toward Intermixing-Free Interfaces. (2022) doi:10.1002/admi.202201997.

22. Longo, E. *et al.* Spin-Charge Conversion in Fe/Au/$Sb_2Te_3$ Heterostructures as Probed By Spin Pumping Ferromagnetic Resonance. *Adv Mater Interfaces* **2101244**, 2101244 (2021).

23. Kumar, A. *et al.* Large-Area MOVPE Growth of Topological Insulator $Bi_2Te_3$ Epitaxial Layers on i-Si(111). *Cryst Growth Des* (2021) doi:10.1021/acs.cgd.1c00328.

24. Pereira, V. M., Wu, C. N., Tjeng, L. H. & Altendorf, S. G. Modulation of surface states in $Sb_2Te_3$/ $Bi_2Te_3$ topological insulator heterostructures: The crucial role of the first adlayers. *Phys Rev Mater* **5**, (2021).

25. Tserkovnyak, Y., Brataas, A. & Bauer, G. E. W. Spin pumping and magnetization dynamics in metallic multilayers. *Phys Rev B Condens Matter Mater Phys* **66**, 1–10 (2002).

26. Celinski, Z., Urquhart, K. B. & Heinrich, B. Using ferromagnetic resonance to measure the magnetic moments of ultrathin films. *J Magn Magn Mater* **166**, 6–26 (1997).

27. Farle, M. Ferromagnetic resonance of ultrathin metallic layers. *Reports on Progress in Physics* **61**, 755–826 (1998).

28. Sun, L., Hao, Y., Chien, C. L., Searson, P. C. & Searson, P. C. Tuning the properties of magnetic nanowires. *IBM Journal of Research and Development* vol. 49 79–102 Preprint at https://doi.org/10.1147/rd.491.0079 (2005).

29. Longo, E. *et al.* Ferromagnetic resonance of Co thin films grown by atomic layer deposition on the Sb2Te3 topological insulator. *J Magn Magn Mater* **509**, 166885 (2020).

30. Montoya, E., McKinnon, T., Zamani, A., Girt, E. & Heinrich, B. Broadband ferromagnetic resonance system and methods for ultrathin magnetic films. *Journal of Magnetism and Magnetic Materials* vol. 356 12–20 Preprint at https://doi.org/10.1016/j.jmmm.2013.12.032 (2014).

31. Tserkovnyak, Y., Brataas, A. & Bauer, G. E. W. Enhanced Gilbert Damping in Thin Ferromagnetic Films. *Phys Rev Lett* **88**, 4 (2002).

32. Maekawa, S., Valenzuela, S. O., Saitoh, E. & Kimura, T. *Spin Current*. *Series on Semiconductor Science and Technology* (Oxford University Press, 2012). doi:10.1093/acprof:oso/9780199600380.001.0001.

33. Harder, M., Gui, Y. & Hu, C. M. Electrical detection of magnetization dynamics via spin rectification effects. *Physics Reports* vol. 661 1–59 Preprint at https://doi.org/10.1016/j.physrep.2016.10.002 (2016).

34. Harder, M., Cao, Z. X., Gui, Y. S., Fan, X. L. & Hu, C. M. Analysis of the line shape of electrically detected ferromagnetic resonance. *Phys Rev B Condens Matter Mater Phys* **84**, 1–12 (2011).

35. Uchida, K. *et al.* Observation of the spin Seebeck effect. *Nature* **455**, 778–781 (2008).

36. Tserkovnyak, Y., Brataas, A., Bauer, G. E. W. & Halperin, B. I. Nonlocal magnetization dynamics in ferromagnetic heterostructures. *Rev Mod Phys* **77**, 1375–1421 (2005).

37. Li, C. H., Van'T Erve, O. M. J., Yan, C., Li, L. & Jonker, B. T. Electrical Detection of Charge-to-spin and Spin-to-Charge Conversion in a Topological Insulator Bi2Te3 Using BN/Al2O3 Hybrid Tunnel Barrier. *Sci Rep* **8**, (2018).

38. Deorani, P. *et al.* Observation of inverse spin Hall effect in bismuth selenide. *Phys Rev B* **90**, 094403 (2014).

39. Dc, M. *et al.* Observation of High Spin-to-Charge Conversion by Sputtered Bismuth Selenide Thin Films at Room Temperature. *Nano Lett* **19**, 4836–4844 (2019).

40. Jamali, M. *et al.* Giant Spin Pumping and Inverse Spin Hall Effect in the Presence of Surface and Bulk Spin-Orbit Coupling of Topological Insulator Bi2Se3. *Nano Lett* **15**, 7126–7132 (2015).

41. Mendes, J. B. S. *et al.* Dirac-surface-state-dominated spin to charge current conversion in the topological insulator (Bi 0.22 Sb 0.78 ) 2 Te 3 films at room temperature. *Phys Rev B* **96**, 1–7 (2017).

42. He, H. *et al.* Enhancement of spin-to-charge conversion efficiency in topological insulators by interface engineering. *APL Mater* **9**, (2021).

43. Rimoldi, M. *et al.* Epitaxial and large area Sb2Te3thin films on silicon by MOCVD. *RSC Adv* **10**, 19936–19942 (2020).

*Supporting Information*

1. **Grazing Incidence X-ray Diffraction on Si(111)/Sb₂Te₃/Bi₂Te₃**

In Figure S1 the GIXRD pattern collected for the Si(111)/Sb₂Te₃/Bi₂Te₃ heterostructure is reported. The measurement is performed by fixing the rocking angle ω at 1.5°. The Miller indices for the Bi₂Te₃ crystalline structures are taken according to the file with the code 74348 of the ICSD database. The high intensity of the (003) and (006) reflections with respect to the powder pattern indicates the high texturization of the film along the [00ℓ] direction, a typical condition for similar chalcogenide-based compounds grown on Si(111) substrates. [1–3] However, the presence of reflections not belonging to the [00ℓ] family of plains indicates a not fully organized crystalline structure, as discussed in the main text.

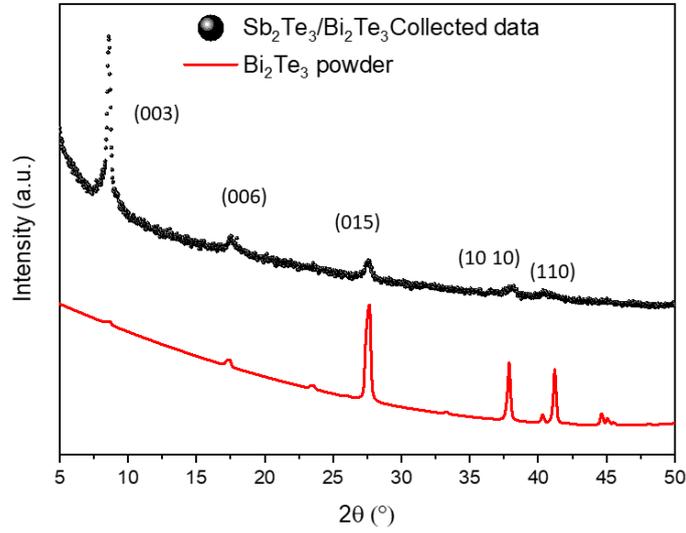

Figure 5: GXRD measurement on the Si(111)/Sb₂Te₃/Bi₂Te₃ heterostructure.

2. **Estimation of the spin mixing conductance and spin current density from FMR experiments**

From BFMR experiments it is possible to extract important information about the magnetization dynamic in a TI/FM system. A fundamental quantity defining the amount of spin current crossing the TI/FM interface is called spin mixing conductance ($g_{eff}^{\uparrow\downarrow}$). To calculate the value of $g_{eff}^{\uparrow\downarrow}$ for heterostructure S1, as reported in the main text, we use the following procedure.

By plotting the $\Delta H(f_{res})$ curve for the heterostructures S0 and S1, the $\alpha$ values for the two heterostructures are extracted by fitting the dataset with the equation:

$$\Delta H = \Delta H_0 + \frac{4\pi}{|\gamma|} \alpha\, f_{res} \qquad (1)$$

According to the SP theory,[4–6] the difference between the damping constant of heterostructures S1 and S0 is proportional to $g_{eff}^{\uparrow\downarrow}$, which can be calculated using the following relation:

$$\alpha_{FM/NM} - \alpha_{FM} = \frac{g\mu_B}{4\pi M_S} g_{\uparrow\downarrow}^{T,eff} \frac{1}{t_{Co}} \quad (2)$$

where $\mu_B$ is the Bohr magneton and the other quantities are defined in the main text. The calculated $g_{eff}^{\uparrow\downarrow}$ is then proportional to the generated 3D spin current density, namely:

$$J_S^{3D} = \frac{\hbar}{4\pi} g_{\uparrow\downarrow}^{T,eff} \left[\vec{M} \times \frac{d\vec{M}}{dt}\right] \quad (3)$$

In a SP-FMR experiment the $J_S^{3D}$ value depends on the measurement conditions, according to the following equation:

$$J_S^{3D} = \frac{Re(g_{eff}^{\uparrow\downarrow})\gamma^2 h_{RF}^2 \hbar}{8\pi\alpha^2} \left(\frac{\mu_0 M_S - \sqrt{(\mu_0 M_S)^2 + 4\omega^2}}{(4\pi M_S \gamma)^2 + 4\omega^2}\right) \frac{2e}{\hbar} \quad (4)$$

where $h_{RF}$ is the strength of the oscillating magnetic field generated by the RF current and $\omega$ the RF frequency. For further details, see Ref.[6].

### 3. Alternative calculation to estimate the SCC efficiency according to the Inverse Spin Hall Effect (ISHE)

For the ISHE the spin current generated into the FM layer is pumped and converted in the bulk states of the spin-sink layer (in our case the $Bi_2Te_3$ layer). Here, the figure of merit of the SCC efficiency is the so-called spin-Hall angle $\theta_{ISHE}^{\lambda_S} = \frac{J_{SP}}{\lambda_S \tanh(\frac{t_{Bi2Te3}}{2\lambda_S}) J_S^{3D}}$, where $\lambda_S$ is the spin diffusion length and $t_{Bi2Te3}$ the thickness of the $Bi_2Te_3$ layer. $J_S^{3D}$ is the spin current density generated in the SP experiment, which varies according to the frequency and the power of the RF current used to excite the system (see above in Supp. Info.) and which is, in our case, $3.28 \cdot 10^5 Am^{-2}$. For a tentative calculation of the $\theta_{ISHE}^{\lambda_S}$, we consider $\lambda_S = 5\ nm$, a reasonable value among those found in literature for chalcogenide-based TIs.[7–10] As a result, we obtain $\theta_{ISHE}^{\lambda_S} = 0.088$, or alternatively a SCC efficiency around the 8.8 %.

### 4. XPS and RHEED measurements of the Si(111)/$Sb_2Te_3$/$Bi_2Te_3$ surface

In order to ensure the proper reconstruction of the $Bi_2Te_3$ surface upon the preparation of the sample to perform ARPES measurements, its chemical and structural evolution are followed by X-ray Photoemission Spectroscopy (XPS) and Reflection High-Energy Electron Diffraction (RHEED) characterization.

The preparation of the sample surface is performed in two steps: firstly, an Ar+ sputtering is carried out at 1.5 KeV and $10^5$ mbar with a duration settled to 150 s, and subsequently an annealing process is performed for a complete reconstruction of the surface. In figure S2 the evolution of the XPS spectra for the unprocessed Si(111)/$Sb_2Te_3$/$Bi_2Te_3$ heterostructure (black line), after the sputtering treatment (red line) and upon the annealing (blue line), are shown. Moreover, to help the surface recovering and to prevent the peeling of the $Sb_2Te_3$ and $Bi_2Te_3$ layers during the annealing, a small amount of Te flow (0.5 Å/s) is delivered to the chamber, being Te the most volatile element.

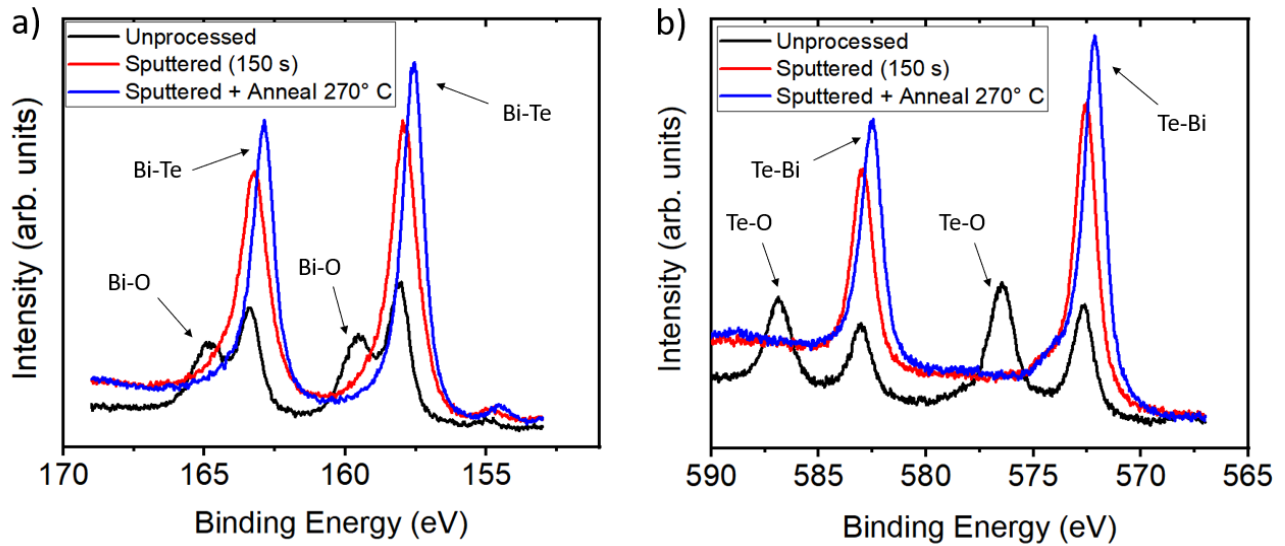

Figure S6:Figure 1: XPS spectra performed on the heterostructure Si/Bi2Te3/Sb2Te3, as it is (black line), after 150 s of Ar+ sputtering (red line) and after an annealing at 270 ° C (blue line). In panel (a) the energy of the 4f orbitals of Bi are shown, while panel (b) displays the 3d orbitals of Te.

In this case, the annealing temperature is decreased with respect to the $Sb_2Te_3$ and $Bi_2Te_3$ single layers studied in Ref.[11], which have been treated at 292 °C. By observing the XPS signals after the annealing (Fig. S2 blue line), it can be seen that the weak shoulder belonging to the Bi-O (panel (a)) and Te-O (panel (b)) bonds disappear, thus confirming the full removal of the oxidized species.

In order to check the evolution of the crystallinity of the Si(111)/$Sb_2Te_3$/$Bi_2Te_3$ surface, RHEED measurements are performed after each treatment. As shown in Fig. S3 (a), RHEED does not reveal any electronics reflection pattern for the unprocessed sample, thus indicating that a continuous native oxide layer caps the whole surface. Differently, Figure S3 (b) shows a visible RHEED pattern by following the sputtering procedure, which is however still not fully defined, indicating a partial damage of the surface crystalline structure. Finally, after the thermal annealing, the RHEED pattern becomes clearly defined, thus demonstrating an almost full reconstruction of the sample surface, as reported in in Fig. S3 (c).

**Bibliography**

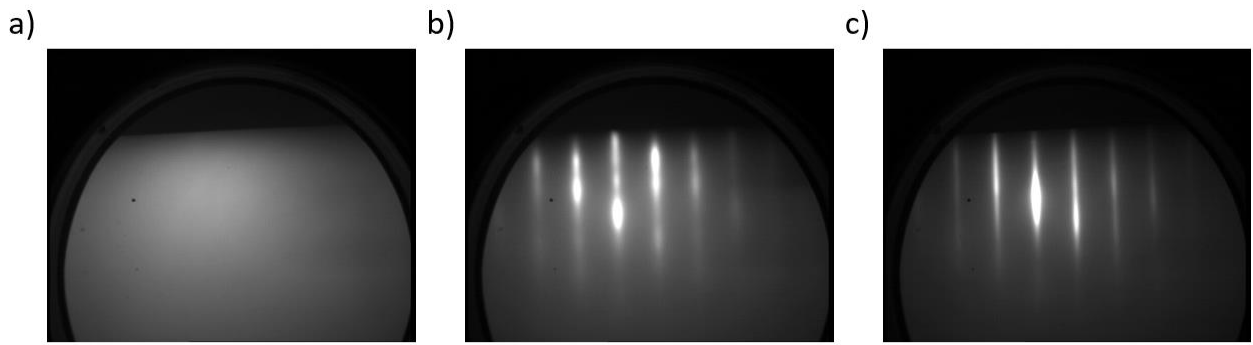

Figure S7: RHEED pattern for various stages of the surface processing: as deposited, in panel (a), after the 150 s Ar+ sputtering, in panel (b) and after sputtering and annealing, panel (c).

1.  Rimoldi, M. *et al.* Epitaxial and large area Sb2Te3thin films on silicon by MOCVD. *RSC Adv* **10**, 19936–19942 (2020).

2.  Rimoldi, M. *et al.* Effect of Substrates and Thermal Treatments on Metalorganic Chemical Vapor Deposition-Grown $Sb_2Te_3$ Thin Films. *Cryst Growth Des* **21**, 5135–5144.

3.  Kumar, A. *et al.* Large-Area MOVPE Growth of Topological Insulator Bi 2 Te 3 Epitaxial Layers on i-Si(111) . *Cryst Growth Des* (2021) doi:10.1021/acs.cgd.1c00328.

4.  Tserkovnyak, Y., Brataas, A. & Bauer, G. E. W. Spin pumping and magnetization dynamics in metallic multilayers. *Phys Rev B Condens Matter Mater Phys* **66**, 1–10 (2002).

5.  Tserkovnyak, Y., Brataas, A. & Bauer, G. E. W. Enhanced Gilbert Damping in Thin Ferromagnetic Films. *Phys Rev Lett* **88**, 4 (2002).

6.  Longo, E. *et al.* Large Spin-to-Charge Conversion at Room Temperature in Extended Epitaxial Sb2Te3 Topological Insulator Chemically Grown on Silicon. *Adv Funct Mater* **2109361**, (2021).

7.  Jamali, M. *et al.* Giant Spin Pumping and Inverse Spin Hall Effect in the Presence of Surface and Bulk Spin-Orbit Coupling of Topological Insulator Bi2Se3. *Nano Lett* **15**, 7126–7132 (2015).

8.  Dc, M. *et al.* Room-temperature high spin–orbit torque due to quantum confinement in sputtered BixSe(1–x) films. *Nat Mater* **17**, 800–807 (2018).

9.  Dc, M. *et al.* Observation of High Spin-to-Charge Conversion by Sputtered Bismuth Selenide Thin Films at Room Temperature. *Nano Lett* **19**, 4836–4844 (2019).

10. Dc, M. *et al.* Room-temperature spin-to-charge conversion in sputtered bismuth selenide thin films via spin pumping from yttrium iron garnet. *Appl Phys Lett* **114**, (2019).

11. Locatelli, L., Kumar, A., Tsipas, P., Dimoulas, A. & Longo, E. Magnetotransport and ARPES studies of the topological insulators  grown by MOCVD on large - area Si substrates. *Sci Rep* 1–10 (2022) doi:10.1038/s41598-022-07496-7.